\documentclass[9pt,twocolumn,twoside]{pnas-old}

\templatetype{pnasresearcharticle} 
\usepackage{graphicx}

\begin{document}

\title{Machine-learning heat flux closure for multi-moment fluid modeling of nonlinear Landau damping}

\author[a]{Ziyu Huang}
\author[a,1]{Chuanfei Dong}
\author[a]{Liang Wang}

\affil[a]{Center for Space Physics and Department of Astronomy, Boston University, Boston, MA 02215}

\leadauthor{Huang}

\significancestatement{
It is generally believed that nonlinear Landau damping can only be captured by kinetic plasma models evolving phase space dynamics. Yet, these models are notably computation-intensive, creating a significant challenge for scaling both in space and time. Here, we propose an innovative approach that integrates kinetic physics from the first-principles Vlasov simulation data into a multi-moment fluid model using machine learning. This new fluid model can successfully reproduce the nonlinear evolution of the Landau damping process from kinetic simulations. We demonstrated that the trade-off between high computational cost in fully kinetic models and the lack of kinetic physics in fluid models can be effectively addressed using machine-learning techniques.
}
 
\authorcontributions{Z.H., C.D., and L.W. designed research; performed research; analyzed data; and wrote the paper.}
\authordeclaration{The authors declare no conflict of interest.}

\correspondingauthor{\textsuperscript{1}To whom correspondence should be addressed. Email: dcfy@bu.edu}


\begin{abstract}

Nonlinear plasma physics problems are usually simulated through comprehensive modeling of phase space. The extreme computational cost of such simulations has motivated the development of multi-moment fluid models. However, a major challenge has been finding a suitable fluid closure for these fluid models. Recent developments in physics-informed machine learning have led to a renewed interest in constructing accurate fluid closure terms. In this study, we take an approach that integrates kinetic physics from the first-principles Vlasov simulations into a fluid model (through the heat flux closure term) using the Fourier neural operator—a neural network architecture. Without resolving the phase space dynamics, this new fluid model is capable of capturing the nonlinear evolution of the Landau damping process that exactly matches the Vlasov simulation results. This machine learning–assisted new approach provides a computationally affordable framework that surpasses previous fluid models in accurately modeling the kinetic evolution of complex plasma systems.

\end{abstract}

\dates{This manuscript was compiled on \today}
\doi{\url{www.pnas.org/doi/10.1073/pnas.2419073122}}

\maketitle
\thispagestyle{firststyle}
\ifthenelse{\boolean{shortarticle}}{\ifthenelse{\boolean{singlecolumn}}{\abscontentformatted}{\abscontent}}{}

%\firstpage[1]{1}

\dropcap{T}he balance between capturing kinetic physics and the modeling computational cost is a long-lasting issue in computational plasma physics and astrophysics. Fully kinetic simulations of plasmas, such as solving Vlasov equations, offer an intricate understanding of microscale plasma physics. Their primary strength lies in accurately modeling the evolution of the distribution function in phase space -- a critical aspect for understanding the kinetic aspect of the problem. However, the high-dimensional nature of these simulations, encompassing both spatial and velocity space, leads to substantial computational costs. This challenge is particularly pronounced when addressing multiscale physics problems such as magnetic reconnection and turbulence \cite{Drake2006,Daughton2011,Fiuza2020,Ji2022,Dong2018,Dong2022,Yang2023,Li2023,Zhou2023} as well as global magnetospheric modeling of planets, pulsars and black holes \cite{WangL2018,Dong2019,Palmroth2023,Philippov2022,Crinquand2022}. Alternatively, fluid models offer a more computationally affordable approach by solving the macroscopic fluid quantities. In comparison to the Vlasov model, fluid models evolve a finite number of fluid moment equations (such as number density, velocity, and pressure) constructed by taking velocity moments of the Vlasov equation. It is generally believed that fluid models fall short in fully encapsulating the comprehensive kinetic features detailed by the Vlasov model.

Recently, several multi-moment fluid models have been developed to partially incorporate electron kinetic physics into a fluid approach \cite{Hakim2008,Wang2015,Huang2019,Allmann-Rahn2018,Wang2020,Ng2020}. The challenge of such fluid models is finding a closure term to approximate the effects of microscopic kinetics on the macroscopic fluid behavior. When integrating the Vlasov equation, the advection term $v \frac{\partial f}{\partial x}$ always brings higher-order moment terms for the low-order moment equations. Consequently, the update of the lower-order moment equations (e.g., the pressure equation) depends on the evolution of the next-higher-order moment
(e.g., the heat flux $q$); therefore the fluid model must include a closure relation to close the system of equations. The reliability of the fluid model is thus influenced by the closure approach, which approximates the moment hierarchy by estimating higher-order moments using their lower-order counterparts. Traditional models of fluid closure using linear or quasi-linear assumptions often fall short in accurately representing the complex and nonlinear behavior of particle and wave within the plasma, particularly in terms of their energy distribution and the impact of these distributions on wave damping and the overall energy balance of the system.

The heat flux closure introduced by Hammett and Perkins (HP) \cite{hammett1990fluid} in 1990 has been recognized as a profound advancement in the field. The HP closure writes
\begin{equation}
\tilde{q}_{k}=-n_{\rm 0}\chi \frac{2^{1/2}v_{t}}{{|k|}} ik\tilde{T}_{k},
\label{HPclosure}
\end{equation}
where $\chi$ is a dimensionless coefficient, $v_{t}$ is the electron thermal speed, $\tilde{T}_{k}=\tilde{p}-T_{\rm 0}\tilde{n}/n_{\rm 0}$ is the Fourier transformation of the perturbed temperature. The HP closure is derived by matching the exact linear response associated with Landau damping in a collisionless electrostatic plasma. It is noteworthy that the HP closure is presented in Fourier space and thus is a non-local closure. Although the fluid model with HP closure shows promising results for various problems \cite{Ng2017}, its numerical integration into fluid codes is highly challenging and expensive, mainly due to its non-local nature. As a result, despite continuous efforts and progress over the years \cite{Hunana2019}, the closure problem applicable to fluid models still stands as a significant research topic yet to be thoroughly addressed and resolved.

Machine learning has appeared as a unique tool in this aspect, capable of lending significant insights \cite{Alves2022,Donaghy2023,Joglekar2023}. Following the HP closure discussed earlier, studies \cite{ma2020machine,wang2020deep} employed several neural networks including multilayer perceptron (MLP) to create a surrogate model for a Landau fluid closure in the configuration space. Furthermore, Multiple Partial Differential Equations Network (mPDE-Net) was used to accurately learn the HP closure which captures the linear Landau damping process with a correct damping rate \cite{cheng2023data}. Many such studies have demonstrated the potential of using machine learning techniques to effectively approximate the heat flux closure.

However, the majority of previous research have been focusing on replicating the HP closure mentioned above which is derived from linear assumptions and frequently referred to solve linear Landau damping.  One exception is the study conducted by \cite{qin2023data}, which employed the physics informed neural networks (PINNs) and gradient-enhanced physics-informed neural networks (gPINNs).  Their model learns an implicit fluid closure from the Vlasov simulation data with a relatively large initial perturbation, which is expected to reach the nonlinear stage during the evolution. However, both PINN  and gPINN in \cite{qin2023data} are only able to predict the linear part of the damping precisely for $t < 10\omega_{pe}^{-1}$ and failed to predict any nonlinear features. Up to now, there have been very few studies addressing the heat flux closure that can capture the nonlinear Landau damping process through either theoretical approaches or machine learning techniques.

Nevertheless, the HP closure provides significant insights towards a general heat flux closure for fluid models. The fact that the HP closure is constructed in Fourier space \cite{hammett1990fluid} motivates us to look for solutions in Fourier space as well. A recently developed machine learning technique called Fourier Neural Operator (FNO) \cite{li2020fourier} shares the same insights for the general purpose of learning partial differential operators. The way that the neural network build connection between different datasets is to identify edges and difference of different parts of the dataset using convolution kernel. However, the smoothing by higher-order differentiations blurs the boundary and edges which naturally brings challenges for convolution kernel. Instead of constructing neural network in real space, FNO builds the neural network in the Fourier domain and identifies operators that link different variables.

Fourier Neural Operators (FNOs) have garnered significant attention in the sphere of numerical modeling and alternative neural-based computations due to their exceptional versatility and generalizing ability over vast multi-dimensional dataset. One of the distinctive strengths of FNO lies in their ability to utilize the Fourier transform to transfer complex data into the frequency domain, enabling neural networks to learn and approximate nonlinear operations by spotting underlying patterns among Fourier coefficients. This unique ability of FNO provides them with the capability to decipher wide-ranging rules for a broad array of PDEs. Further advancements like U-FNO \cite{wen2022u} and implicit FNO \cite{you2022learning} have further refined this methodology, gradually growing in their ability to address more complex problems.

In this study, we take an innovative approach by developing a heat flux closure for a multi-moment fluid model using FNO. The benefit of a neural operator is to explore unknown equations that capture relations of different variables which naturally fit the puzzles of finding closure for the multi-moment fluid model. This machine-learning heat flux closure is numerically integrated into a fluid model that successfully reproduces the nonlinear Landau damping (as well as linear Landau damping) from kinetic simulations.

\section{Model Setup}

The architecture of our machine learning model, including inputs and outputs, 
is illustrated in Fig. \ref{fig:Model}. We use the Fourier Neural Operator (FNO) to map the lower-order fluid moments (e.g., density $n$, velocity $u$, and pressure $p$) to the heat-flux gradient $\partial q / \partial x$. Fig. 1 consists of three principal steps: (\emph{i}) an initial linear layer $P$ that lifts the input $\mathbf{a}(x)$ to a higher-dimensional latent space, 
(\emph{ii}) a series of four Fourier layers (\texttt{FL}) that extract global and local features, and (\emph{iii}) a final linear layer $Q$ projecting the latent representation back to the physical dimension of the output $\mathbf{u}(x)$.

In each Fourier layer, the data follow two parallel paths. 
Along the \emph{upper path}, we apply a Fourier transform, 
a channel-wise linear transformation in the frequency domain, 
and then an inverse Fourier transform. 
Along the \emph{lower path}, we apply a simple linear operator $W$ in real space, 
which helps capture non-periodic and localized features. 
The outputs from these two paths are added elementwise 
and passed through a ReLU activation function. 
Symbolically, if $\mathbf{a}^\ell(x)$ denotes the input to layer $\ell$, then the upper path is
\[
\boldmath{\mu}^\ell = \mathcal{F}^{-1}
\Bigl(
  {R} \bigl[\mathcal{F}\{\mathbf{a}^\ell(x)\}\bigr]
\Bigr),
\]
and the lower path is
\[
\boldmath{\nu}^\ell = W\bigl(\mathbf{a}^\ell(x)\bigr),
\]
where $\mathcal{F}$ and $\mathcal{F}^{-1}$ represent the forward and inverse Fourier transforms, 
${R}$ is a learnable linear operator in Fourier space, 
and $W$ is a learnable linear operator in real space. 
The sum of $\boldmath{\mu}^\ell$ and  $\boldmath{\nu}^\ell$ is finally passed through a nonlinear activation function 
to produce $\mathbf{a}^{\ell+1}(x)$. 
This structure leverages both the global spectral information 
(via the Fourier transforms) 
and local real-space information (via $W$).

Our goal is to establish a general mapping from lower order moments, such as $\{n(x),u(x),p(x)\}_{t = t_i}$ at any given time step $t_i$, and anticipate ${\frac{\partial q(x)}{\partial x}}_{t = t_i}$. We also test the results using individual moments such as $n(x)$, $u(x)$ and $p(x)$, but the performance does not meet our expectations. This is primarily because when one or more of these moments are nearly constant in the configuration space at some time step, their features in Fourier space amount to random noise. The combination of these three moments considerably lowers the chances of such issues. We also note that a preceding study \cite{laperre2022identification} on machine learning closure followed a similar strategy for selecting inputs.

We employ the open source code GKEYLL that includes a Vlasov solver \cite{juno2018discontinuous} to generate training and testing datasets. Here we study the 1-D Landau damping problem as Landau damping is one of the most fundamental kinetic processes in collisionless plasmas. The first-principles modeling of collisionless Landau damping is described by Vlasov equation:

\begin{equation}
\frac{\partial f_{\mathrm{s}}}{\partial t}+\mathbf{v_s} \cdot \nabla_{\mathbf{r}} f_{s}+\left(\frac{e_{s}}{m_{s}} \right)\mathbf{E} \cdot \nabla_{\mathbf{v}} f_{s}=0
\end{equation}
where $f_{s}(\mathbf{r}, \mathbf{v_s}, t)$ is the phase space distribution function of particle species $\rm{s}$, ${e_s}/{m_s}$ is the charge to mass ratio, and $\mathbf{E}$ is the electric field that is updated through the Maxwell equations.

The fluid quantities $n_s$, $u_s$, $p_s$ and $q_s$ of species $s$ can be calculated  as follows:

\begin{equation}
\begin{gathered}
n_s(x, t)=\int f_s\left(x, v, t\right) d v, \\
u_s(x, t)=\frac{1}{n_s(x, t)} \int v_s f_s\left(x, v, t\right) d v \\
p_s(x, t)=m_s \int\left(v_s-u_s\right)^2 f_s\left(x, v, t\right) d v \\
q_s(x, t)=m_s \int\left(v_s-u_s\right)^3 f_s \left(x, v, t\right) d v .
\end{gathered}
\label{Eq:moments}
\end{equation}

The ions are treated as an immobile and neutralizing ion background. The initial perturbation for electron species, $e$, is configured as follows:

\begin{equation}
n_e(x, t=0)=n_0\left(1+A \cos \left(k x\right)\right) \text {, }
\end{equation}
\begin{equation}
n_i(x, t=0)=n_0,
\end{equation}
where $n_0$ is the initial number density, $A$ and $k$ are the amplitude and wavenumber of the perturbation, respectively. 

We choose the parameter of $k=0.35 \lambda_{D e}^{-1}$ and the perturbation amplitude $A=0.1$  that is large enough to initiate the nonlinear damping in long term evolution. The simulation domain of our study is periodic, where $0<x<{2 \pi}/{k}$, segmented into 64 sections. The velocity space ranges from $-6 v_{t h_s}$ to $6 v_{t h_s}$, and is similarly divided into 64 segments.

We run the Vlasov simulations until $t=40 \omega_{p e}^{-1}$ so that our dataset includes both the early linear phase 
and the later nonlinear regime where phase-space holes form. We sample the low-order moments $n(x)$, $u(x)$ and $p(x)$ shown in Eq. \ref{Eq:moments} from the Vlasov simulation data as the inputs and the gradient of heat flux $\frac{\partial q(x)}{\partial x}$ as the output. The training dataset is selected with a time step of $\Delta t = 0.005 \,\omega_{pe}^{-1}$. FNO learns the closure relationship with those 8000 steps and predicts the closure for 20000 steps with a smaller and different time step of $\Delta t = 0.002 \,\omega_{pe}^{-1}$. This setup compels the model to generalize beyond the linear stage and accurately capture nonlinear Landau damping.

A comparison of the gradient of heat flux $\frac{\partial q}{\partial x}$ between the Vlasov simulation data and the FNO output is presented in Fig.~\ref{fig:dqdx_comparsion}. The absolute error is decreased to the order of $10^{-4}$ demonstrating that FNO can accurately capture the mapping from $n$, $u$, $p$ to $\frac{\partial q}{\partial x}$. The errors have shown to be evident starting from $t =30 \omega_{pe}^{-1}$, which is the time when the nonlinear Landau damping starts to develop. However, the errors are still within $2 \times 10^{-4}$, indicating that FNO is also able to capture the mapping during the nonlinear stage of the Landau damping process.

It is noteworthy that in this study we use the low-order moments, $n(x)$, $u(x)$ and $p(x)$ from the entire simulation domain as the neural network input, similar to the non-local HP closure (see Eq. \ref{HPclosure}). Though not shown here, we have also attempted to learn local heat flux closures using the deep operator network (DeepONet) \cite{lu2021learning}. For instance, we use $n$, $u$, and $p$ values at ${(x_{j-2}, x_{j-1}, x_j, x_{j+1}, x_{j+2})}$, as input to the neutral network, to predict the heat flux gradient, $\partial q / \partial x$, at the point, $x=x_j$. Promisingly, the neutral network can also accurately capture the correct heat flux gradient, $\frac{\partial q}{\partial x}$, closely resembling the results in Fig. \ref{fig:dqdx_comparsion}. This finding sheds light on discovering an accurate local fluid closure.

\begin{figure*}[bth]

\begin{minipage}{.45\textwidth}
\begin{equation}
\begin{aligned}
& \frac{\partial n}{\partial t}+\frac{\partial}{\partial x}(n u)=0 \\
\vspace{10mm}
&\frac{\partial u}{\partial t}+u \frac{\partial u}{\partial x}+\frac{1}{m_e n} \frac{\partial p}{\partial x}=\frac{q_e}{m_e} E_x,\\
\vspace{10mm}
& \frac{\partial p}{\partial t}+u \frac{\partial p}{\partial x}+3 p \frac{\partial u}{\partial x} = -\frac{\partial q}{\partial x},\\
\vspace{10mm}
& \frac{\partial E_x}{\partial t}+\frac{q_e}{\varepsilon_0} n u=0
\end{aligned}
\label{Eq:Tem-moment}
\end{equation}
\end{minipage}%
\hfill
\vline
\hfill
\begin{minipage}{.45\textwidth}
\begin{equation}
\begin{aligned}
\Huge
& \frac{\partial n}{\partial t}+\frac{\partial}{\partial x}(n u)=0 \\
&\frac{\partial u}{\partial t}+u \frac{\partial u}{\partial x}+\frac{1}{m_e n} \frac{\partial p}{\partial x}=\frac{q_e}{m_e} E_x, \\
& \frac{\partial p}{\partial t}+u \frac{\partial p}{\partial x}+3 p \frac{\partial u}{\partial x}= 
\begin{gathered}
{\includegraphics[width=0.1\columnwidth]{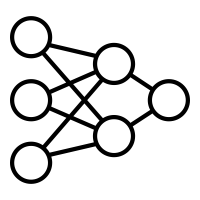}}
\end{gathered}\\
& \frac{\partial E_x}{\partial t}+\frac{q_e}{\varepsilon_0} n u=0\\
\end{aligned}
\label{Eq:Tem-moment-ML}
\end{equation}
\end{minipage}
\end{figure*}

\begin{figure*}
\centering
\includegraphics[width=1.0\linewidth]
{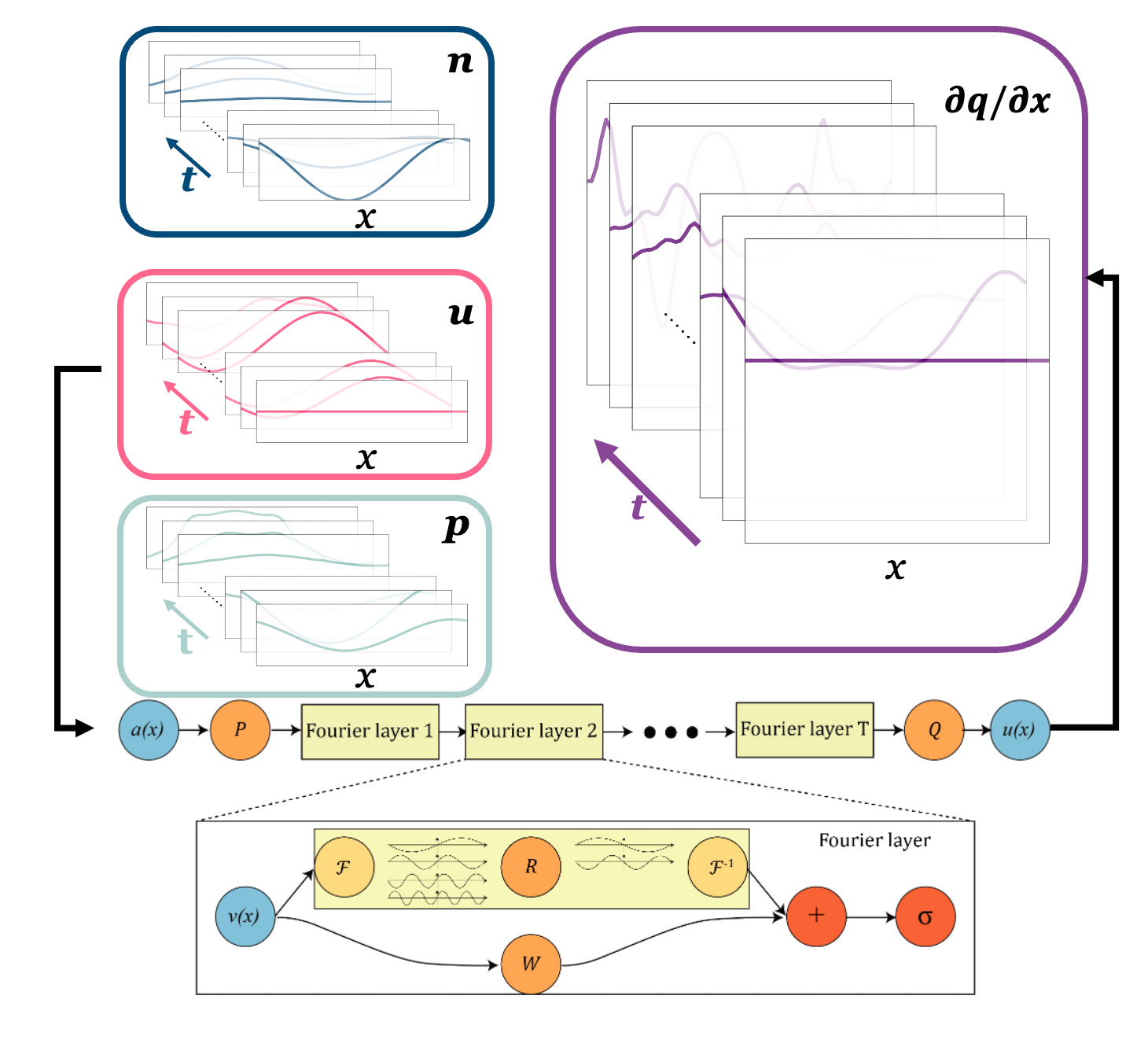}
\caption{Architecture of the Fourier Neural Operator (FNO) for predicting the heat flux gradient, \( \partial q / \partial x \), from low-order moments (\( n, u, p \)). The input moments are lifted to a higher-dimensional latent space using an initial fully connected layer (\( P \)), processed through four successive Fourier layers, and projected back to the physical dimension via a final fully connected layer (\( Q \)). The figure illustrates how the Fourier layers split the data into two paths: (1) the \emph{upper path} applies a Fourier transform, channel-wise linear transform, and inverse Fourier transform; and (2) the \emph{low path} applies a learnable linear operator in real space for problems with non-periodic boundary conditions. The outputs of both paths are combined and passed through a ReLU activation.}
\label{fig:Model}
\end{figure*}

\begin{figure*}
\centering
\includegraphics[width=0.8\linewidth]{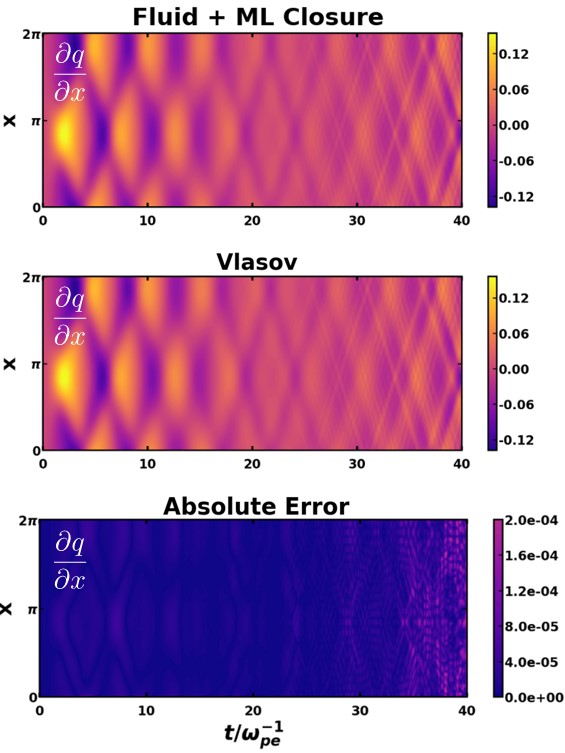}
\caption{Comparison of the heat flux gradient, \( \partial q / \partial x \), from Vlasov simulations and the Fourier Neural Operator (FNO). \emph{Top} panel: Heat flux gradient calculated directly from Vlasov simulations, serving as the ground truth. \emph{Middle} panel: Heat flux gradient predicted by FNO, trained using kinetic simulation data. \emph{Bottom} panel: Absolute difference between the Vlasov simulation results and FNO predictions.}
\label{fig:dqdx_comparsion}
\end{figure*}

Our objective is to numerically integrate kinetic effects into a multi-moment fluid model through machine learning. The newly developed fluid model will strike a balance between computational practicality and a superior degree of physical accuracy. Eq. \ref{Eq:Tem-moment} are the multi-moment fluid equations derived by taking moments of the Vlasov equation. The term $\frac{\partial q}{\partial x}$ is the gradient of the heat flux which relies on the fluid closure model such as the HP closure \cite{hammett1990fluid}. To achieve our objective, the term $\frac{\partial q}{ \partial x}$ is replaced by a neural network (see Eq. \ref{Eq:Tem-moment-ML}), indicating $\frac{\partial q}{ \partial x}$ is calculated at each time step by FNO. We employ a fourth-order Runge--Kutta (RK4) scheme for the time integration of the multi-moment fluid model. At $t=0$, we initialize all the fluid moments including $n$, $u$, $p$, $q$ and electric field $E_x$. During the very first time step, $n$, $u$, $p$ and $E_x$ are updated using the prescribed initial $q$ in the fluid equations. For each subsequent time step, we pass the newly updated lower-order moments $\{n, u, p\}$ to our trained FNO, which provides an approximation of the heat-flux gradient $\left(\partial q/\partial x\right)_{\mathrm{FNO}}$. This value is then used within the RK4 solver to feed back into Eq. \ref{Eq:Tem-moment-ML} and update $n$, $u$, $p$ and $E_x$.

The neural network acts as a source term providing accurate heat flux gradient through gathering the features of lower order moments in Fourier space, similar to HP closure but without following linear assumptions. It is noteworthy that we will learn the PDEs of the Vlasov-Amp\'{e}re system in this study given that the Vlasov-Amp\'{e}re and Vlasov-Poisson systems are equivalent if the initial conditions of the Vlasov-Amp\'{e}re system satisfy Poisson's equation \cite{Xie2013}.

\section{Numerical Results}

\begin{figure*}
\centering
\includegraphics[width=1.0\linewidth]{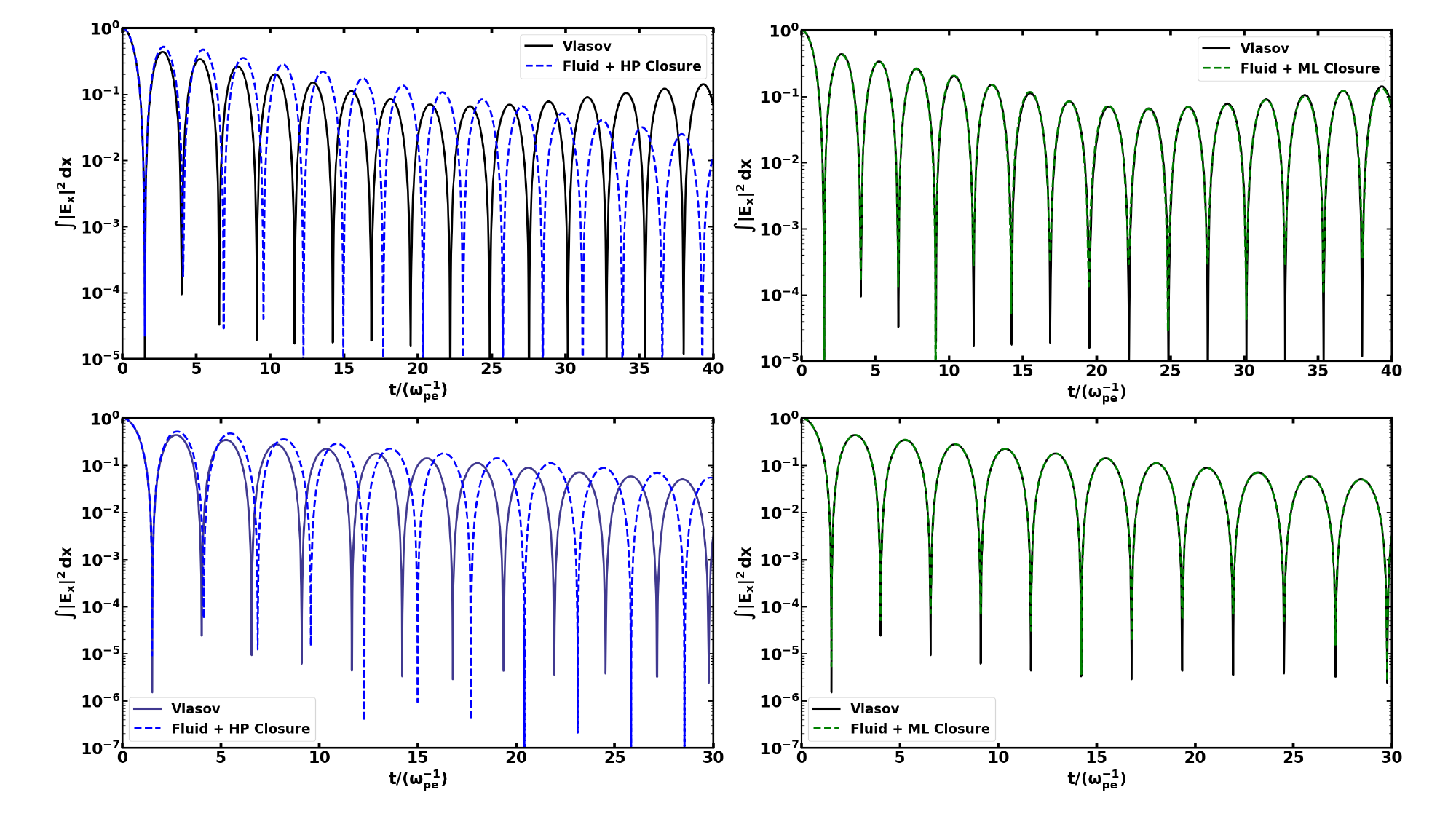}
\caption{Normalized electric field energy comparison between the Fluid + HP model (\emph{Left} column) and Fluid + ML model (\emph{Right} column) against Vlasov simulation data. The \emph{Top} panels depict a case of nonlinear Landau damping, while the \emph{Bottom} panels correspond to a linear damping case.}
\label{fig:E_energy}
\end{figure*}

\begin{figure*}
\centering
\includegraphics[width=1.0\linewidth]{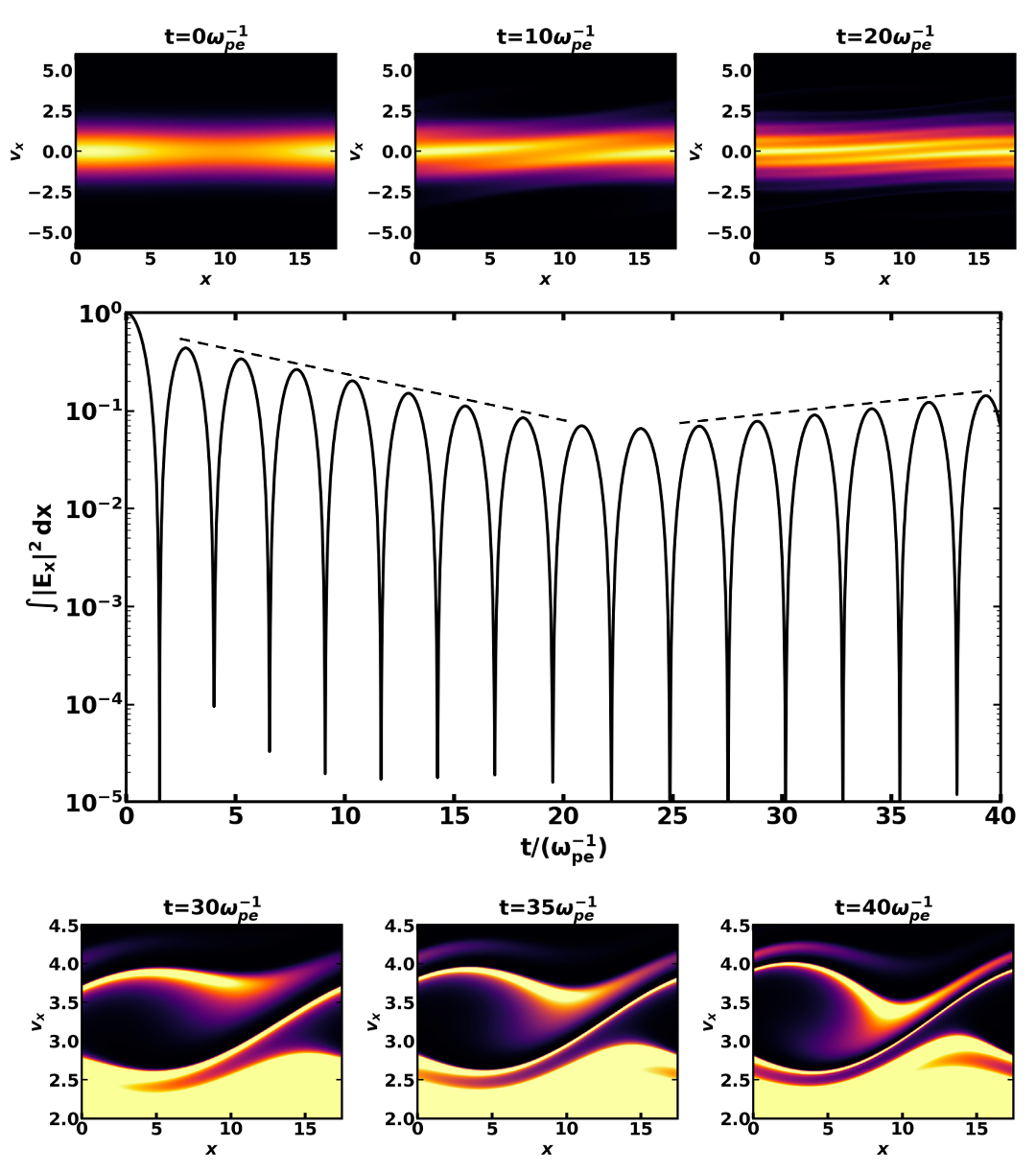}
\caption{Time evolution of the electric field energy and corresponding phase-space diagrams during linear and nonlinear stages of Landau damping. Linear stage snapshots are shown for \( t = 0, 10, 20 \, \omega_{\mathrm{pe}}^{-1} \), while nonlinear stage snapshots are at \( t = 30, 35, 40 \, \omega_{\mathrm{pe}}^{-1} \). The electric field energy evolution provides insight into the transition from linear to nonlinear stages of Landau damping, and the phase-space diagrams illustrate particle trapping at the nonlinear stage.}
\label{fig:Gkeyll_phase_heatflux}
\end{figure*}

Fig. \ref{fig:E_energy} depicts the evolution of the electric field energy predicted by the multi-moment fluid model with the machine learning heat flux closure and with the HP closure, denoted by ``Fluid + ML" and the latter denoted as ``Fluid + HP", respectively. 
Upon examining the evolution of electric field energy in the top panel of Fig. \ref{fig:E_energy}, we notice the decay of the electric field energy at a rate of $\gamma=-0.1493$ during the initial stages. Subsequently, the damping ceases and it transitions to the growth phase with $\gamma=0.0396$ during the nonlinear evolution stage. The Vlasov simulations provide the phase space plot of the distribution function which allows us to understand the evolution of the Landau damping process at different stages. Snapshots of the phase space plots at the damping stage are shown in the top panels of Fig. \ref{fig:Gkeyll_phase_heatflux} which shows the phase mixing in the linear stage. The phase space hole starts to form around $t=25 \omega_{pe}^{-1}$ where the damping ceases. The bottom panels of Fig. \ref{fig:Gkeyll_phase_heatflux} depict the complicated phase space structures in the nonlinear evolutionary stage. Surprisingly, the ``Fluid + ML" model is able to reproduce the evolution of the electric field energy from the Vlasov simulations. In contrast, the ``Fluid + HP" model falls short in resolving the growth phase with a constant damping rate of $\gamma=-0.1257$.

In addition to resolving the nonlinearity, the ``Fluid + ML" model also excels in capturing the right bouncing frequency. As shown in the top left panel of Fig. \ref{fig:E_energy}, the two curves start to deviate after two bouncing periods due to the different bouncing frequencies. This is made evident in Fig. \ref{fig:FNO_nupex_error}, where we plot the absolute error of the two model outputs compared to the Vlasov model $\left\|y_{\text {true }}^i-y^i\right\|$. Meanwhile, we also test a case with a smaller perturbation, where $A = 0.05$ akin to the case investigated by \cite{cheng2023data}, to focus on the linear Landau damping process in the bottom panel of Fig. \ref{fig:E_energy}. Again, the ``Fluid + ML" model can reproduce the Vlasov simulation data, demonstrating the robustness of the trained heat flux closure. Although the ``Fluid + HP" cannot capture the right bouncing  frequency, it gives the right damping rate.

\begin{figure*}
		\centering
		\includegraphics[width=1.0\linewidth]{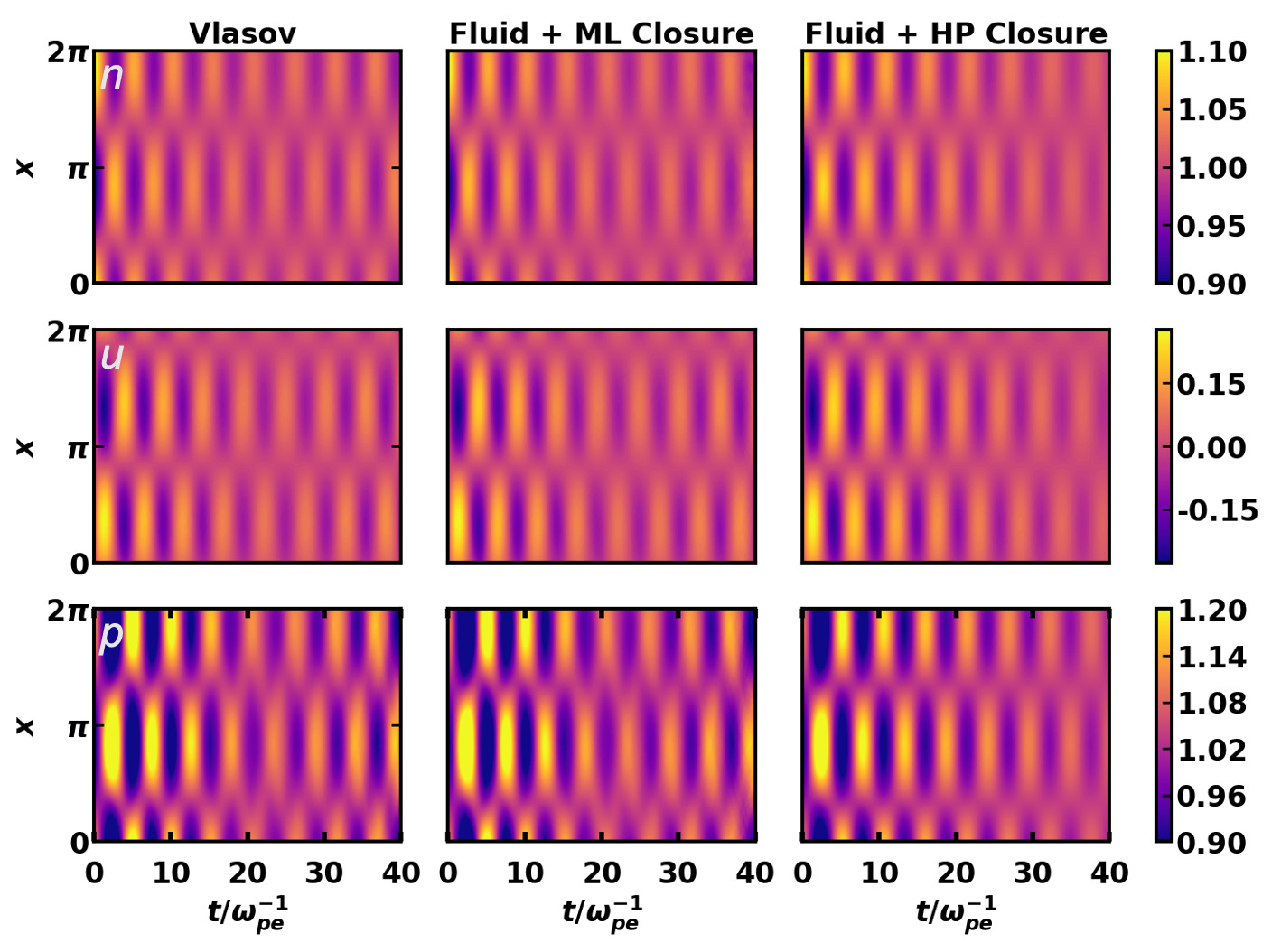}
		\caption{Comparison of fluid moments (\( n \), \( u \), \( p \)) from the Vlasov simulation model, the Fluid + ML model, and the Fluid + HP model. Each row displays the number density (\( n \)), fluid velocity (\( u \)), and pressure (\( p \)) obtained from each model. In the third row, the Vlasov simulation shows pressure values beginning to increase starting from \( t = 20 \, \omega_{pe}^{-1} \). The Fluid + ML model closely follows this trend, accurately capturing the pressure increase, whereas the Fluid + HP model fails to capture the pressure increase.}
		\label{fig:FNO_nupex}
\end{figure*}

\begin{figure*}
		\centering
		\includegraphics[width=1.0\linewidth]{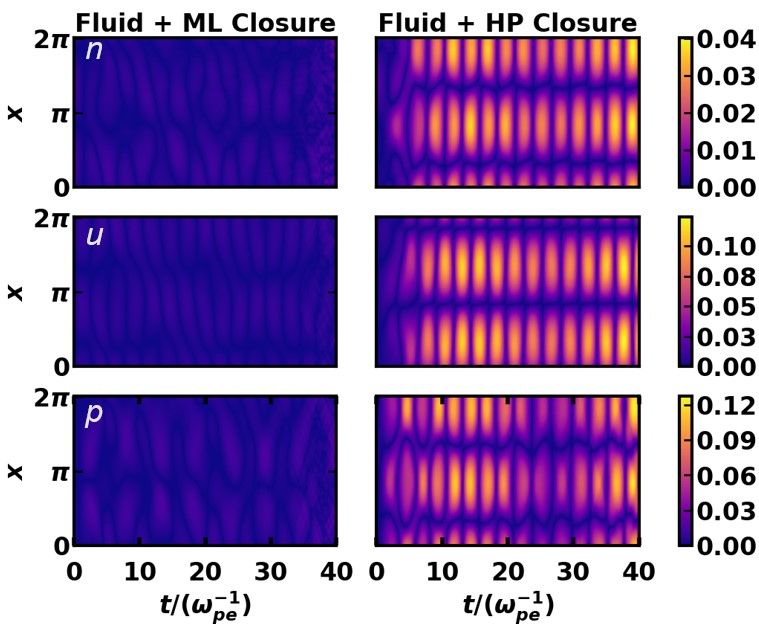}
		\caption{Absolute error of fluid moments (\( n \), \( u \), \( p \)) for the Fluid + ML model (first column) and Fluid + HP model (second column) with respect to the Vlasov simulation data. Each row represents the absolute errors for number density (\( n \)), bulk velocity (\( u \)), and pressure (\( p \)), respectively. Notably, the Fluid + HP model exhibits a significant phase  discrepancy in the time evolution, whereas the Fluid + ML model closely matches the Vlasov simulation results.}
		\label{fig:FNO_nupex_error}
\end{figure*}

The evolution of lower order moments, simulated by the Vlasov model, the ``Fluid + ML" model, and the ``Fluid + HP" model, are displayed in Fig. \ref{fig:FNO_nupex}. From top to bottom, we compare the number density, velocity, and pressure among three models. Generally, the ``Fluid + ML" model demonstrates superior performance compared to the ``Fluid + HP" model, especially starting from approximately $t = 25 \omega_{pe}^{-1}$, when the system evolves from linear to nonlinear stage as indicated in Fig. \ref{fig:Gkeyll_phase_heatflux}. From the bottom panel of Fig. \ref{fig:FNO_nupex}, it is clear that the ``Fluid + ML" model accurately reproduces all the patterns from the Vlasov simulations, whereas the ``Fluid + HP" model exhibits damping trend only and fails to capture the right bouncing frequency. A comparison of the absolute error is presented in Fig. \ref{fig:FNO_nupex_error}, where the ``Fluid + ML" model produces significantly smaller errors compared to the ``Fluid + HP" model.

\section{Conclusions}

The integration of kinetic effects in a fluid model through machine learning by providing an accurate fluid closure has proven to be promising. The fluid model is able to capture certain kinetic physics that is important when modeling space and astrophysical plasmas. In this work, we study the Landau damping process by coupling a multi-moment fluid model with a FNO network architecture, where FNO provides the fluid closure as a source term in the pressure equation. We train the FNO for the heat flux closure using the simulaiton data from the first-principles Vlasov model which include all the kinetic physics.

In comparison of the ``Fluid + ML" model with the ``Fluid + HP" model, we find that the former significantly outperforms when modeling both the linear and nonlinear stages of the Landau damping process and capturing the right bouncing frequency. For the first time, we are able to capture the nonlinear Landau damping process using a fluid model even after the phase space holes form. This breakthrough implies that the machine learning technique exhibits a higher degree of adaptability and accuracy when dealing with complex dynamics. 

This study illustrates a category of reduced models that can match the performance of full models. The full model in this context refers to the Vlasov model which is computationally expensive, limiting its broader application in space, astrophysical, and laboratory plasma physics. However, leveraging machine learning has enabled reduced fluid models to produce promising results at lower computational costs. 

It is noteworthy that the trained neural networks and the associated heat flux closure learned in this study may be generalized across various nonlinear plasma processes by utilizing transfer learning. Transfer learning enables a foundation model trained on one specific plasma phenomenon to be adapted to other related phenomena with minimal additional training data. Therefore, transfer learning will enable the development of a comprehensive nonlinear closure framework for diverse plasma processes.

The general idea of integration of machine learning techniques to reduced models sheds light on the accurate and efficient modeling of large-scale systems, which can be extended to complex multiscale laboratory, space, and astrophysical physics problems. In the future, we plan to apply this approach to higher-dimensional problems, such as magnetic reconnection, a ubiquitous process in magnetized plasmas throughout the Universe. \\

\showmatmethods{} % Display the Materials and Methods section

\noindent{\textbf{Data, Materials, and Software Availability.} The training datasets from Vlasov simulations have been deposited in Zenodo (\url{https://doi.org/10.5281/zenodo.14915239}). The datasets contain two data files for the case of linear and nonlinear Landau damping, respectively.}

\acknow{The authors thank H. Fu and Y. Qin for the helpful discussions. This work was supported by NASA grant 80NSSC23K0908 and the Alfred P. Sloan Research Fellowship. Resources for this work were provided by the NASA High-End Computing Program through the NASA Advanced Super-computing Division at Ames Research Center. We also would like to acknowledge high-performance computing support from National Energy Research Scientific Computing Center, a DOE Office of Science user facility, and from Derecho provided by NCAR’s CISL, sponsored by NSF. For distribution of the model results used in this study, please contact the corresponding author.}

\showacknow{} % Display the acknowledgments section

\section{Reference}
% Bibliography
%\bibliography{pnas-sample}

%\newpage

%\newpage

\newpage

\newpage

\newpage

\newpage

\end{document}